\documentclass[twocolumn,preprintnumbers,amsmath,amssymb,pre,showkeys]{revtex4}

\usepackage{txfonts}
\usepackage{amsfonts}
\usepackage{amsmath}
\usepackage{graphicx}
\usepackage{dcolumn}
\usepackage{bm}
\usepackage{overpic}
\usepackage{multirow}
\usepackage{float}
\usepackage{soul}
\usepackage{color}

\usepackage{hyperref}
\hypersetup{colorlinks={true},allcolors={blue}}  

\begin{document}

\title{Localized energy absorbers in Hertzian chains\footnote{doi:~\href{https://doi.org/10.1103/PhysRevApplied.18.014078}{10.1103/PhysRevApplied.18.014078}}}
\email[Corresponding author: ]{carlos.vasconcellos@usach.cl}
\author{Carlos Vasconcellos$^{1\ast}$}
\author{Ren{\'e} Zu{\~n}iga$^1$}
\author{St{\'e}phane Job$^2$}
\author{Francisco Melo$^1$}

\affiliation{$^1$Department of Physics of Universidad de Santiago de Chile, Avenida Ecuador 3493, 9170124 Estaci\'on Central, Santiago, Chile and Center for Soft Matter Research, SMAT-C, Avenida Bernardo O'Higgins 3363, Estaci\'on Central, Santiago (Chile)}
\affiliation{$^2$Laboratoire Quartz, EA-7393, ISAE-Supm{\'e}ca, 3 rue Fernand Hainaut 93400 Saint-Ouen-sur-Seine (France)}

\date{\today}

\begin{abstract}
Energy absorbers and energy-harvesting devices have been under the scope of scientists and engineers for decades to fulfill specific technological needs, mainly concerned with sound and vibration absorbers, and efficient mechanical energy converters. 
In this paper, as a proof of concept, we build a mass-in-mass device to study the response of a linear absorber immersed in one of the spheres composing a linear array of equal elastic spheres. Spheres barely touch one another and can thus sustain nonlinear solitary wave propagation only. The linear intruder absorbs a given amount of energy depending on the frequency content of the incident solitary wave. A numerical simulation is developed to account for the experimental finding. The validation of the numerical model allows for the theoretical study of the energy absorbed by any number of intruders, and to demonstrate that the former increases exponentially with the latter, indicating that only ten of the intruders is enough to absorb the system energy. A detailed study of the transmitted energy from an external source into the chain reveals that, due to nonlinearity, the array of spheres is able to convert almost any mechanical shock to a well defined solitary or trains of solitary waves, whose frequency content is nearly independent on the excitation amplitude. This property leads to the design of a device, which is optimized to absorb energy over a broad frequency range.
\end{abstract}


\maketitle

%==========%
\section{\label{sec:introduction}INTRODUCTION}

Mass-in-mass (MIM) systems~\cite{Huang2009a,Huang2009b,Huang2010,Finocchio2014,Lydon2014,Tan2014,Bonanomi2015,Dertimanis2016,Kevrekidis2016,An2017,Banerjee2017,Wallen2017,Banerjee2018,Xu2019,Zivieri2019,Faver2020,Fiore2020,Song2021} and nonlinear energy sinks (NESs)~\cite{Gendelman2001,Vakakis2001,Vakakis2009,Vakakis2018} are promising strategies for designing tailored metamaterials with unusual acoustical and mechanical properties. MIM lattices have been investigated for more than a decade~\cite{Huang2009a,Huang2009b,Huang2010}, leading to stimulating concepts and applications. For example, equivalent models to represent a lattice system consisting of MIM units have been introduced and the need for a negative-mass-density concept demonstrated~\cite{Huang2009a}. Wave attenuation properties and energy-transfer mechanisms of metamaterials, modeled by MIM lattices, have been captured through negative-effective-mass density~\cite{Huang2009b}, which has been found also useful in the description of dispersion curves and band-gap structures of multiresonators~\cite{Huang2010}.
Periodic graded one-dimensional (1D) metacomposites for enlarging the band-gaps~\cite{An2017}, and
frequency-graded 1D metamaterials have also been envisaged~\cite{Banerjee2017}. For instance, a potential solution towards low frequency and wideband acoustic or vibration insulation has been proposed through the design of tuned frequency-graded arrangement of resonating units capable of extending the attenuation band of 1D metamaterial~\cite{Banerjee2018}.

Applications of locally resonant MIM in the domain of seismic isolation have been recently proposed~\cite{Dertimanis2016, Fiore2020}, and the impact of nonlinear harmonics on the attenuation mechanisms identified~\cite{Fiore2020}. Hierarchically organized local resonators, which possess the ability to efficiently tailor elastic wave or vibration attenuation to various frequency regions~\cite{Xu2019}, are among contemporary strategies of designing metamaterials. Active control of strongly nonlinear periodic systems has recently been explored, leading to effective manipulation of band structures of the lattice~\cite{Song2021}.
The localized modes of vibration involved in MIM systems are valuable alternatives for implementing sound metamaterials due to their exceptional properties in the nonlinear regime~\cite{Lydon2014, Bonanomi2015}. Indeed, in contrast to localized modes arising in a mechanical lattice due to mass or rigidity defects, resonant defects can be tuned from being extremely localized to totally delocalized by an external force~\cite{Lydon2014}. The possibility of inducing coupling in a network of resonant defects through the tuning of the localization distance appears as a possible strategy to control energy transmission, which may be useful for the design of specific switches for sound transmission.
A classical way to enhance energy dissipation, for the control of mechanical vibrations and wave attenuation, is to consider a viscoelastic material (like a polymer or rubber) together with a resonant system: the loss factor of a mass-spring-dashpot system is indeed $\eta=c/2\sqrt{km}$, it is thus possible to tune the attenuation (i.e. how fast and how much potential and kinetic energy can be dissipated) by appropriately choosing a stiffness $k$ and/or a mass $m$ at given constant viscous coefficient $c$. A well-known practical example is the tuned mass damper (TMD)~\cite{DenHartog1934,Sadek1997,Rana1998,Krenk2005,Lee2006,Hoang2008,Krenk2014}, which takes advantage of an amplification of the mechanical response, at the resonant frequency $\omega=\sqrt{k/m}$, to enhance the frequency-dependent dissipation in a narrow tunable spectral band. 
In the same spirit, locally resonant metamaterials can exhibit a higher dissipation throughout the spectrum, generally refereed to as metadamping~\cite{Hussein2013, Brunet2013, Manimala2014,Frazier2015,Chen2016,Bacquet2018,DePauw2018,Li2019,Abbasi2020}.
The practical implementation of metadampers~\cite{Hussein2013,Manimala2014,Frazier2015}, relies on the inclusion of distributed locally resonant scatterers in a surrounding elastic matrix~\cite{Brunet2013}. In the long wavelength approximation, the local resonators lead to an effective medium exhibiting an artificial dissipative feature that can be engineered at will, like in a TMD. In this sense, MIM phononic lattices thus result to be the discrete counterparts of the metadampers.
The main weakness of linear resonant systems relies on the fact that they are frequency selective: broadband excitations do not fit. Instead, NES~\cite{Gendelman2001,Vakakis2001,Vakakis2009} are well-known nonlinear counterparts of the TMD that can accommodate such a limitation. They include a nonlinear component (generally the stiffness), which provokes an irreversible and broadband targeted energy transfer (TET) across spatial or temporal scales~\cite{Vakakis2018}. These systems are proven to be useful when implemented to control linear primary systems; their nonlinear nature provides a passive self-tunable capability to adapt the frequency response to the strength of an excitation. 
In this frame, arrays of highly nonlinear elements have been found to exhibit remarkable capabilities to broaden sharp impact and vice versa~\cite{Job2005,Melo2006,Job2007b,Job2008,Job2009,Theocharis2009}. For example, this phenomenon is found in 1D metacomposites of spheres barely touching one another in both the stepped~\cite{Job2007b} and tapered chain configurations~\cite{Melo2006}. A solitary wave propagating through the large-diameter section array, at the boundary with the array section of small diameter, splits in several solitary waves, which are narrower and propagate faster. A sequence of solitary wave, of decreasing amplitude and width, is generated. Thus, generation of solitary waves provides a mechanism of transfer of energy from low to high frequency, depending on the ratio of sphere diameter. 

Here, our paper deals with a system taking advantage of a succession of compact resonators embedded in a nonlinear phononic lattice made of an alignment of nonresonant spherical particles: such a lattice supports solitary wave with compact spatial and time support. This means that the waves propagating along these lattices have a well-defined frequency content that can be tuned to match the frequency of the inclusions. This contrasts noticeably with NES (a nonlinear damper embedded in a linear system): our system is a nonlinear lattice inducing a TET while a linear resonant damper further dissipates irreversibly the mechanical energy. It is thus possible to adapt any excitation to the frequency response of the damped linear resonator.
As a proof of concept, we build a mass-in-mass device to study the response of a linear absorber immersed in one of the spheres composing a linear array of equal elastic spheres. Spheres barely touch one another and can thus sustain nonlinear solitary wave propagation. The linear intruder absorbs a given amount of energy depending on the frequency content of the incident solitary wave. A numerical simulation is developed to account for the experimental finding. The validation of the numerical calculation allows for the theoretical study of the energy absorbed by any number of intruders, and to demonstrate that the absorbed energy varies exponentially with the intruder number, indicating that only a few intruders is enough to absorb the system energy. A detailed study of the transmitted energy from an external source into the chain reveals that, due to nonlinearity, the array of spheres is able to convert almost any energy pulse to a well-defined solitary or trains of solitary waves, whose frequency content is nearly independent on the excitation amplitude. This property leads to the design of a device, which is optimized to absorb energy at a broad band.

The paper is organized as follows: In Sec.~\ref{sec:exp_setup} experimental devices, measurements protocols, and intruder preparation are given. In Sec.~\ref{sec:exp_results} the main features of the solitary wave passage through a sphere holding an in-mass intruder are investigated experimentally. In Sec.~\ref{sec:num_results} numerical simulations are developed and contrasted to experimental results. The intruder energy and the energy transferred due to solitary wave passage are studied. These results are generalized to the case of several spheres holding intruders. The progressive energy transfer from solitary to intruders is investigated. Finally, in Sec.~\ref{sec:conclusions} discussions and conclusions are given. 

%==========%
\section{\label{sec:exp_setup}EXPERIMENTAL SETUP}

\begin{figure}[b]
\centering
\includegraphics[width=\columnwidth]{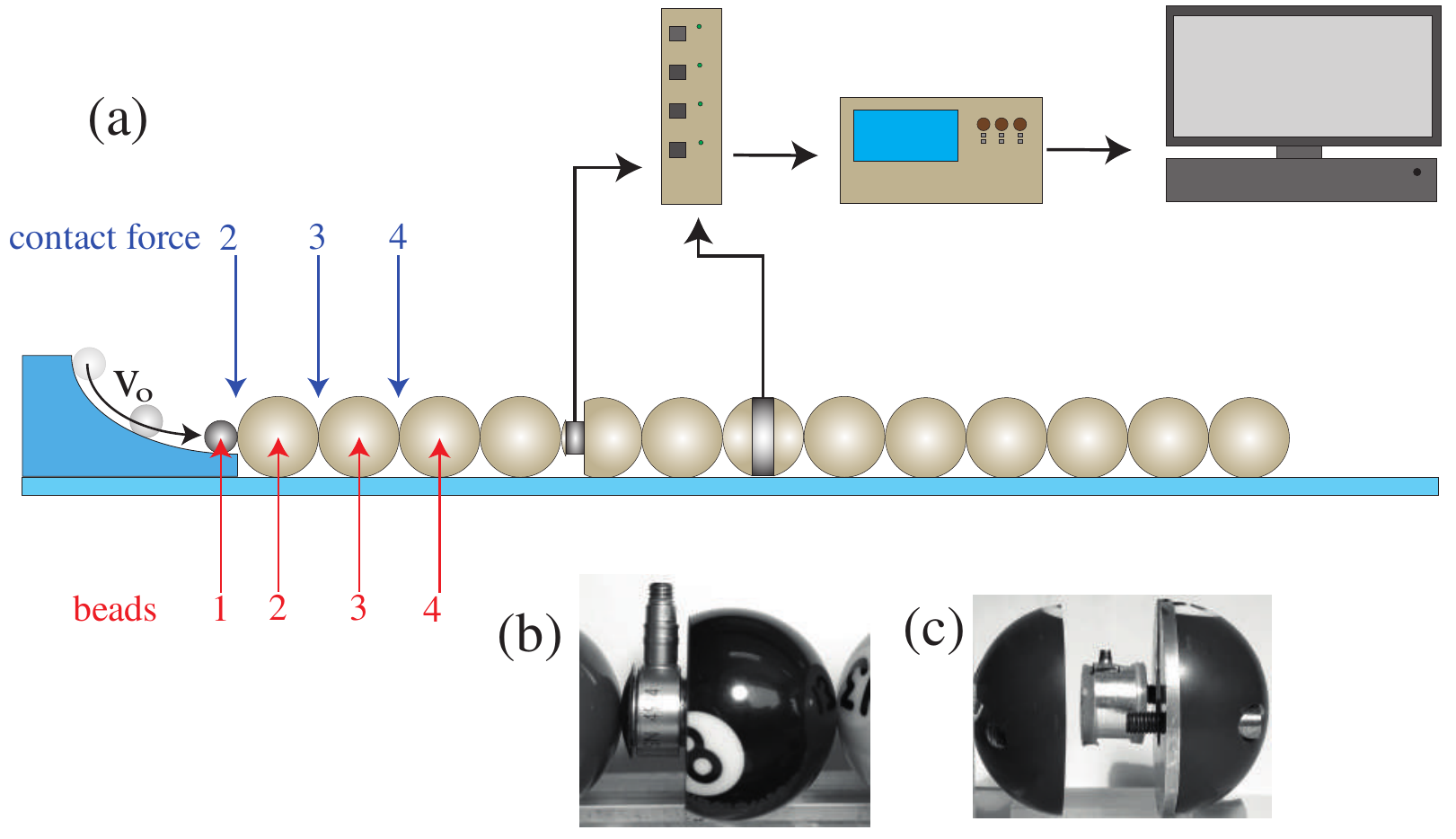}
\caption{Experimental setup. (a) Schematic view showing a linear chain of beads including a bead sensor, a resonant absorber and data acquisition facilities. To excite wave propagation the chain is impacted by a small striker bead. The contact forces are labeled in blue and positions of beads in red. (b) Force sensor embedded in a bead. (c) Energy absorber: a rigid cylindrical accelerometer stands for the internal mass, clamped between two-half spheres via soft polymer disks on each sides, acting as linear springs inside.}
\label{fig:setup}
\end{figure}

Bakelite spheres have been chosen to illustrate the proof of principle, because this material is easy to machine compared to steel beads used in previous experiments. Bakelite has low density $\rho=1775$~kg/m$^{3}$, is nearly incompressible, Poisson ratio $\nu=0.44$, and has a relatively low Young's modulus, $Y=3.9$~GPa, that can accommodate large strain with low plastic deformation. In addition, it is fairly elastic material, i.e. weakly dissipative.
The chain is made of 16 equal beads, with radius $R=19$~mm and mass $M=51$~g, see Fig.~\ref{fig:setup}(a).
The beads, barely touching one another, are aligned on a horizontal Plexiglas track. A nonlinear compressive wave is initiated from the impact of a small striker made of steel, with mass $M_s=3$~g, radius $R_s=4.5$~mm, density $\rho_s=7980$~kg/m$^{3}$, Poisson ratio $\nu_s=0.27$, and Young's modulus, $Y_s=203$~GPa. The striker trajectory is guided by a curved ramp, see Fig.~\ref{fig:setup}(a), which ensures nearly centered impact with an acceptable repeatability. Adjusting the initial height of release allows precise control of the collision velocity of the striker on the first bead of the chain. The pulse is monitored by measuring the load with a piezoelectric transducer (PCB 200B02, with sensitivity $11.24$~mV/N and stiffness $1.9$~kN/$\mu$m) inserted inside a bead cut in two parts, as shown in Fig~\ref{fig:setup}(b). The total mass of the sensor bead matches the mass of a regular bead. The embedded sensor thus allows nonintrusive measurements of the force along the chain.

One of the spheres hosts a resonant inclusion composed of a mass and a linear spring, as shown in Fig.~\ref{fig:setup}(c). The internal mass is a rigid cylindrical accelerometer (PCB 352A24, with sensitivity $10.2$~mV/N and mass $m=3.7$~g) inserted in a cylindrical cavity that is machined in the host bead. The cavity is centered to maintain a nearly homogeneous mass distribution and the host bead is cut into two equal halves to facilitate assembly. The accelerometer is clamped between the two caps via thin disks made of an elastic polymer placed on each side: the whole thus performs as a resonant energy absorber. Here, the accelerometer intrinsically senses its absolute movement, i.e. the acceleration of the internal mass. Note that the host bead includes a metallic ring to tune its total mass to that of a regular bead of the chain. Polymers used are commercial ones (Zhermack-SpA: Elite Double) available in various Young's modulus. These polymer layers have been designed to act as two opposite springs whose spring constant and damping factor can be tuned by changing their thickness and adding dissipative compound to the polymeric matrix. Here, the selected polymer layers have a thickness of about $1$~mm and are made of the Elite Double 32 of Young's modulus $1.11$~MPa, providing an approximate value of the elastic constant of about $k\approx 3.1\times10^{5}$~N/m.
Concerning the data-acquisition facilities, the signals provided by the force sensor and the accelerometer are amplified by a conditioner PCB482A16, recorded by a two-channel numeric oscilloscope Tektronix TDS2012B, and transferred to a computer.

%==========%
\section{\label{sec:exp_results}EXPERIMENTAL RESULTS}

\begin{figure}[b]
\centering
\includegraphics[width=\columnwidth]{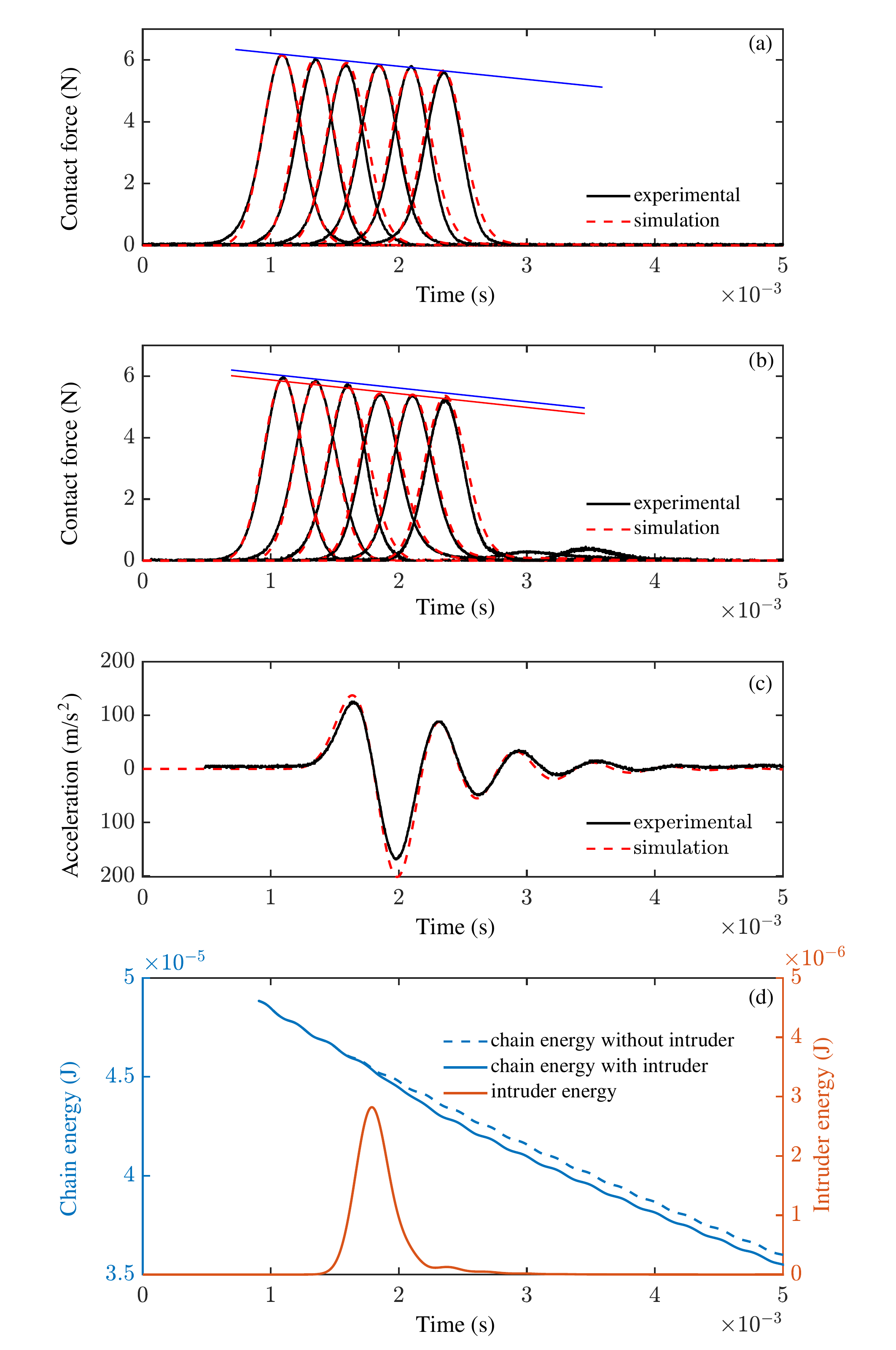}
\caption{Force at the contacts of beads as a solitary wave passes through consecutive bead contacts, obtained one at the time by relocating the sensor bead at successive positions. (a) Force due to solitary wave propagation at contacts 6 to 11. (b) Contact forces in the presence of a bead with the intruder located at position 8. Blue and red lines indicate the additional force decrease experienced by the propagating solitary wave when passing through the host bead. (c) Acceleration signal indicating that energy has been transferred to the intruder during the passage of the solitary wave. For all cases, the solitary wave is excited by a steel striker ($R_s=4.5$~mm and $V_s= 0.45$~m/s). Dashed lines in (a)-(c) obtained through numerical simulations. (d) Numerical calculations of the chain energy in the absence and in the presence of the intruder. A fraction of incident solitary wave energy is transmitted to the intruder. All numerics are performed with the experimental parameters, $R=19$~mm, $\rho=1775$~kg/m$^{3}$, and $\nu=0.44$, and adjusted values $Y=4.5$~GPa and $\tau=4.2$~$\mu$s.}
\label{fig:exp_vs_sim}
\end{figure}

As an experimental benchmark, we first study the solitary wave propagation in the absence of the bead containing the intruder. Solitary waves are excited at one free end of the chain through the impact with the striker, ($R_s=4.5$~mm) at speed, $V_{s}=0.45$~m/s. As in early work~\cite{Job2005}, solitary wave propagation through the chain can be investigated nonintrusively by means of the embedded sensor, see Fig.~\ref{fig:setup}(b), which respects the translational symmetry of the chain by keeping both invariant the contact between spheres and the mass distribution. Thus, the embedded sensor accurately senses the force at the respective contact, which reflects the features of the solitary wave traveling through it. By positioning the force sensor at consecutive locations and recording the force at the respective contacts, the solitary wave evolution through the chain can be investigated. The force amplitude decrease with the distance traveled by the solitary wave, indicating that some energy was dissipated by the lattice, see Fig.~\ref{fig:exp_vs_sim}(a).
The bead with intruder is then introduced in the chain: an additional decrease of the solitary wave amplitude, of a few percent, after crossing the host bead, is observed, see Fig.~\ref{fig:exp_vs_sim}(b), indicating that some energy has been transferred to the intruder. Observation of the acceleration of the intruder confirms that it vibrates at its natural frequency, dissipating its energy after a few oscillations, see Fig.~\ref{fig:exp_vs_sim}(c). 

In order to provide a methodology that allows for the efficient exploration of the main design parameters of the absorber, we first develop a numeric modeling, which includes relevant experimental details. 
Indeed, our numerical simulation based on a Runge-Kutta algorithm allows for the exploration of the main features of solitary waves in interaction with a localized vibration in an alignment of $N$ particles, by solving the nonlinear system of $N$ equations, 

\begin{eqnarray}
M\ddot{U}_{n}&=&\kappa_{n-1,n}( U_{n-1}-U_{n})^{3/2} \nonumber \\
&-&\kappa_{n,n+1}( U_{n}-U_{n+1})^{3/2} \nonumber \\ 
&+&\tau\kappa_{n-1,n}\frac{3}{2}(U_{n-1}-U_{n})^{1/2} (\dot{U}_{n-1}-\dot{U}_{n}) \nonumber \\
&-&\tau\kappa_{n,n+1}\frac{3}{2}(U_{n}-U_{n+1})^{1/2}(\dot{U}_n-\dot{U}_{n+1})\label{eq:wave_equation}
\end{eqnarray}

\noindent for $n=(1\ldots N)$, with $\kappa_{n,n+1}=(\theta_{n} + \theta_{n+1})^{-1}\times(1/R_{n} + 1/R_{n+1})^{-1/2}$ and $\theta_n=3(1-\nu_{n}^2)/(4Y_n)$, being $U_n(t)$, $R_n$, $Y_n$, and $\nu_n$, the instantaneous displacement, the radius, the Young's modulus, and the Poisson coefficient, respectively, of particle $n$. We approximate the main mechanism responsible for the dissipation as an internal viscoelastic behavior of the bulk material, which can be expressed as, $F_D=\tau\times \partial_{t}F_H$, with $F_H=\kappa\delta^{3/2}$~\cite{Kuwabara1987} the Hertzian elastic force, $\delta$ the overlap deformation, and $\tau$ the relaxation time associated to the viscous damping of Bakelite spheres.

An intruder is designed as a linear harmonic oscillator with a linear dissipation. If a particle contains an intruder we replace and solve the equation of the concerned degree-of-freedom $n$ in Eq.~\ref{eq:wave_equation} by the coupled system of equations,

\begin{eqnarray}
(M-m)\ddot{U}_{n}&=&k(u_{n}-U_{n})+c(\dot{u}_{n}-\dot{U}_{n}) + F_c,\\
m\ddot{u}_{n}&=&k(U_{n}-u_{n})+c(\dot{U}_{n}-\dot{u}_{n}),
\end{eqnarray}

\noindent where $F_c$, representing the contact forces acting on the left and right side of the bead, is equal to the right-hand side of the Eq.~\ref{eq:wave_equation}, where $u_n(t)$ is the instantaneous displacement of the inner intruder, and where $k$ and $c$ are the elastic constant and the viscous friction coefficient, respectively. Our calculations accurately reproduce the solitary wave propagation and attenuation, see Fig.~\ref{fig:exp_vs_sim}(a)-(c), by adjusting the relaxation time of spheres and the Young's modulus, obtaining $\tau=4.2$~$\mu$s and $Y=4.5$~GPa, respectively. Moreover, the behavior of intruder is fairly well captured by using the experimental values, for all parameters, by tuning the value of spring constant and by adjusting the viscous dissipation due to the polymeric springs. We obtain $k=3.95$~N/m and $c=12$~N~s/m. As a check of consistency, the frequency response of the intruder has been experimentally characterized by sweeping the excitation frequency in the range of 100 to 2500~Hz, see Fig.~\ref{fig:sm_fig_1} of the Appendix. Fitting the frequency response with a degree-of-freedom resonator, see Sec.~\ref{sec:appendix_1} of the Appendix, indicates that the resonant frequency and the damping coefficient are $f_{res}=\sqrt{k/m}/2\pi \approx $1600~Hz and $c \approx 12 $~N~s/m, respectively.  Both these values are consistent with the ones obtained {\em in situ}, from the fit of the wave experiment mentioned above.  Moreover, the resonant frequency is also compatible with the value obtained from the intruder mass and the spring constant calculated using the parameters of the polymer layer holding the intruder, although the latter is more sensitive to uncertainties.

Thus, our numeric procedure is validated, which provides us with an accurate tool to explore the dependence of energy absorption on parameters that are difficult to access experimentally. However, the value of Young's modulus of Bakelite obtained through the adjusting procedure is about $15$\% higher than the nominal value, which is quantitatively compatible with a stiffening effect due to the thwarted rotations of the particles by the friction, as discussed in Ref.~\cite{Job2005}. The value of the viscoelastic relaxation time, also confirms that Bakelite is a weakly dissipative material: an estimation of the loss factor $\eta=\tan{(\omega\tau)}$ indicates $\eta\simeq10^{-2}\ll1$ since $\omega\tau\simeq(\partial_{t}F_H/F_H)\tau\simeq\tau/\tau_{SW}$ with $\tau_{SW}\simeq0.5$~ms the typical half-duration of the solitary wave pulse, see Fig.~\ref{fig:exp_vs_sim}(a) for instance.

%==========%
\section{\label{sec:num_results}NUMERICAL RESULTS}

%----------%
\subsection{Local energy}

In order to study how energy is transferred and dissipated to the local oscillator, we define the energy of sphere $n$ as, 
\begin{equation}
E_{n}=\frac{1}{2}M\Dot{U}_{n}^{2} + \frac{2}{5}\kappa_{n-1,n}(U_{n-1} - U_{n})^{5/2}
\end{equation}
and the energy of the intruder writes,
\begin{equation}
E_{n}^{int}= \frac{1}{2}m\Dot{u}_{n}^{2}+ \frac{1}{2} k (u_{n} -U_{n})^{2}.
\end{equation}

The total chain energy ($E_T=\Sigma E_n$) evolution in time can be calculated numerically and the case with intruder compared to that without intruder, see Fig.~\ref{fig:exp_vs_sim}(d). The presence of the intruder is revealed by the sudden decrease (nearly 4\%) of the chain energy observed when the solitary wave passes through the host bead, see Fig.~\ref{fig:exp_vs_sim}(d). Energy transfer and energy dissipation occur at nearly the same time scale, see the solid red line in Fig.~\ref{fig:exp_vs_sim}(d).

\begin{figure}[b]
\centering
\includegraphics[width=\columnwidth]{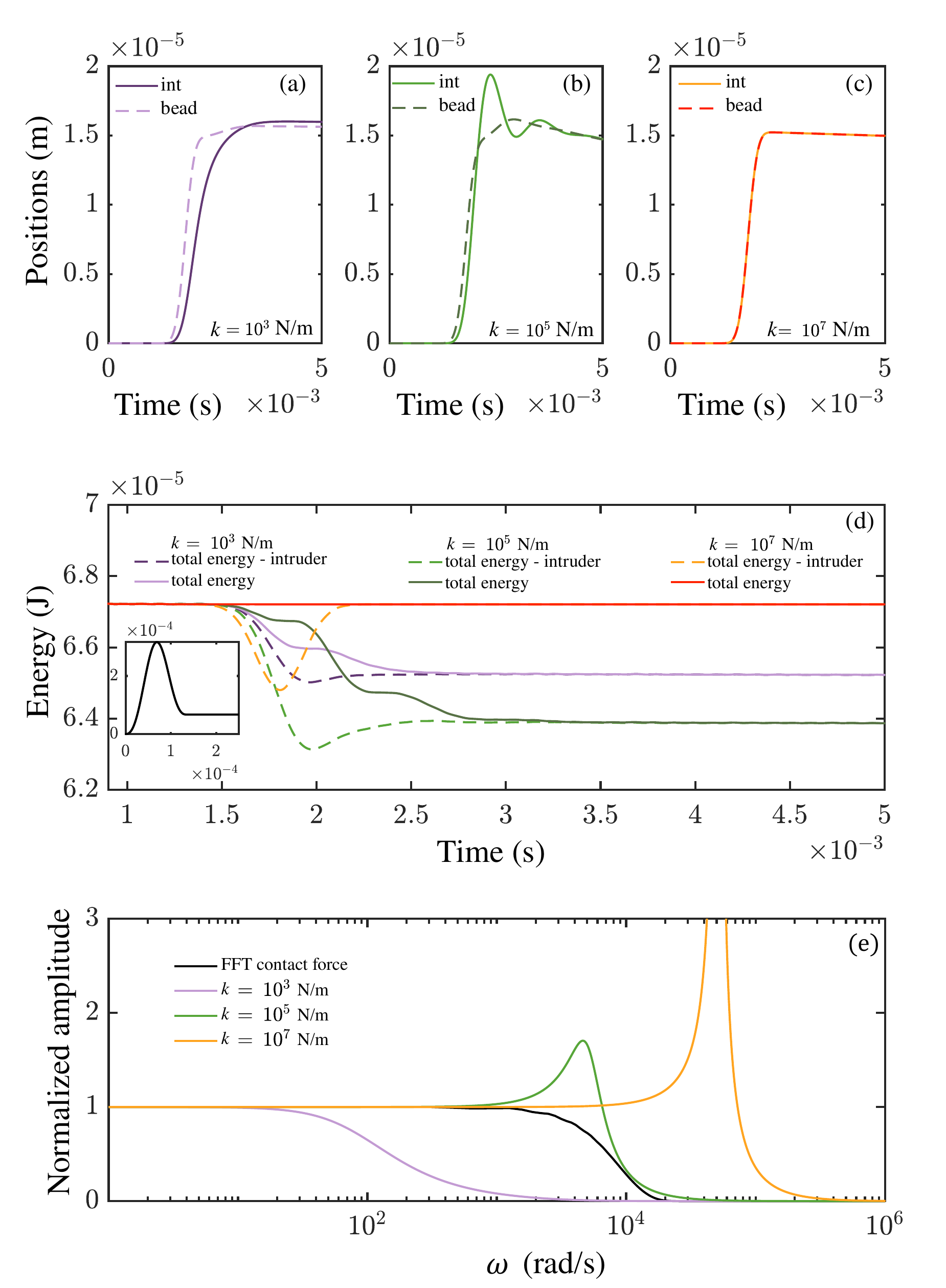}
\caption{(a-c) Motion of both the host bead and the intruder, for three distinct values of the spring constant $k$ of the intruder, and constant viscous coefficient, $c=12$~N~s/m for equal excitation of the chain (steel striker, $R_s=4.5$~mm at $V_s=0.45$~m/s). (a) $k=10^3$~N/m, (b) $k=10^5$~N/m, (c) $k=10^7$~N/m. (d) Total chain energy evolution at early stages of solitary wave intruder interaction, for the same values of intruder stiffness as discussed in (a-c). Inset: injected energy into the chain. A fraction of the energy of the incident striker pass through the chain. In all cases, the host bead is located at position 8 and the chain composed of 38 bead. (e) Intruder frequency response for distinct spring constants and frequency content of the incident solitary wave (solid black line).}
\label{fig:absorcion}
\end{figure}

\begin{figure*}[ht]
\centering
\includegraphics[width=\textwidth]{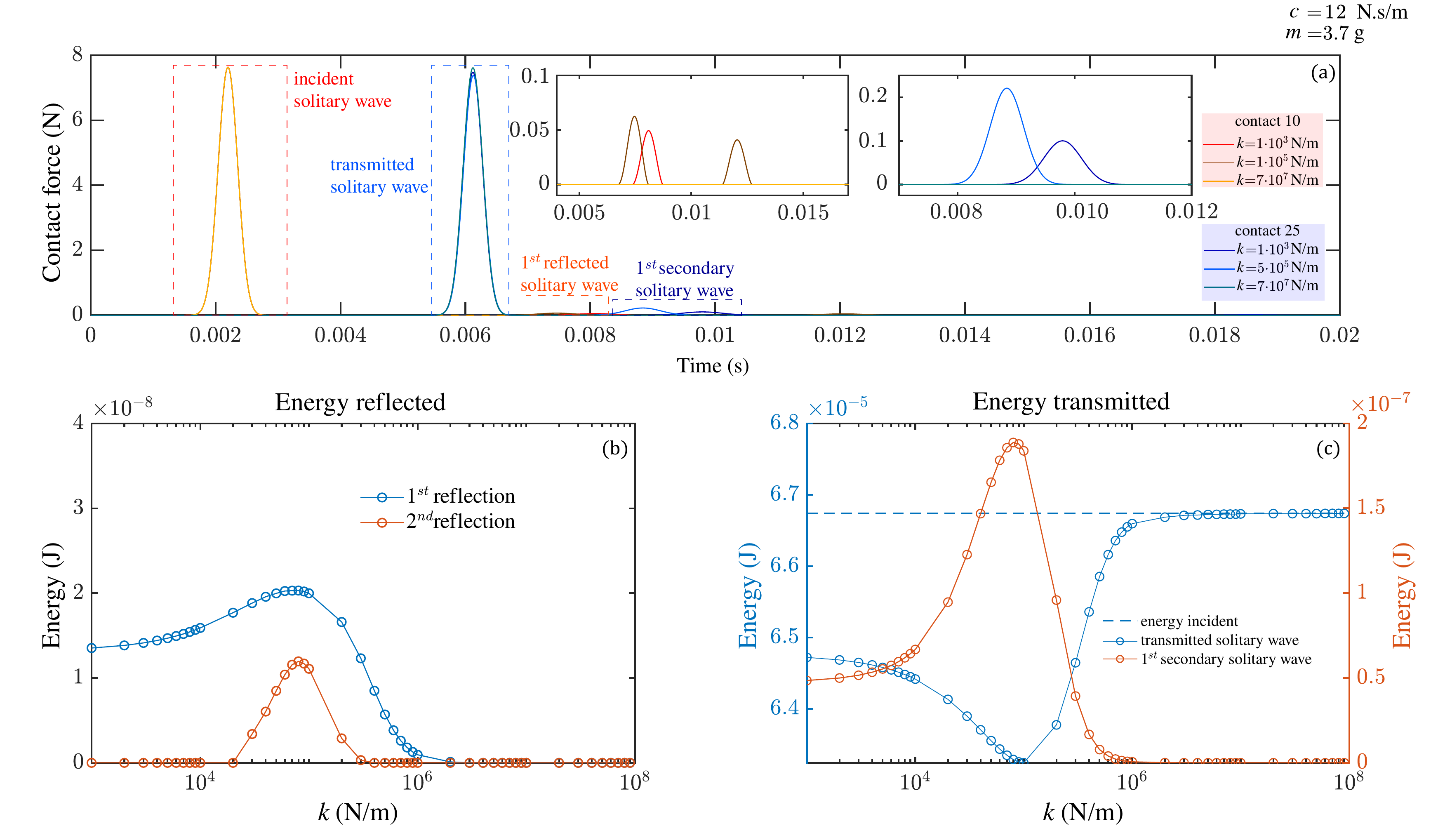}
\caption{ (a) Forces at contacts 10 and 25, illustrating main and secondary waves after interaction with the oscillator, which is located at sphere 15, for three values of $k$. Left inset: reflected secondary waves. Right inset: transmitted secondary waves. (b) Mechanical energy associated with reflected secondary waves for distinct values of $k$. (c) Mechanical energy associated with transmitted secondary waves for the same values of $k$. For all cases, the chain is composed of 46 spheres, $c=12$~N~s/m and $m=3.7$~g.}
\label{fig:all_parameter}
\end{figure*}

%----------%
\subsection{Transfer of energy}

In this section we investigate how the transfer of energy depends on the resonance frequency of the local oscillator. In practice this can be done by changing the spring constant by simply replacing the stiffness of the polymeric layers. In order to isolate the effect of the intruder stiffness, we perform this analysis numerically, neglecting the dissipation of the beads, but we consider that the rate of energy dissipated by the intruder is close to the experimental value, $c\approx 12$~N~s/m. For soft springs [$k=10^3$~N/m, see Fig.~\ref{fig:absorcion}(a)], the intruder motion is similar to that of the host bead but some delay occurs; for very high spring constant [$k=10^7$~N/m, see Fig.~\ref{fig:absorcion}(c)], the intruder moves rigidly with its host; for intermediate spring constant [$k=10^5$~N/m, see Fig.~\ref{fig:absorcion}(b)], the relative displacement of the intruder is clearly enhanced, see the early overshoot. The effect of the enhancement of the relative motion affects the energy absorbed by the intruder, see Fig.~\ref{fig:absorcion}(d). At very high spring constant the intruder just moves with the host and therefore it acquires some kinetic energy. Since no relative motion occurs between the intruder and the host bead, the kinetic energy of the intruder is transferred back to the solitary wave without dissipation. However, at intermediate spring constant, significant relative motion occurs, leading to large deformation of the viscoelastic elastomer caps. Only a very small amount of the energy captured by the intruder is thus transferred back to the transmitted solitary wave, the rest being dissipated by the viscoelasticity. Finally, at low spring constant, the intruder motion is over damped. Here, the energy transfer is less efficient because of the large impedance mismatch between the elastic surrounding medium and the highly viscous intruder, leading to a large reflection of the wave before it gets dissipated (see below). These observations indicate that the conditions to achieve optimized energy transfer to the oscillator depend on both the intruder frequency response and the frequency content of the incident solitary wave. Indeed, maximum overlap of both spectra leads to better energy transfer, as depicted in Fig.~\ref{fig:absorcion}(e). 

To explore which parameters optimize the energy absorption of the oscillator, we vary both the inner spring constant and the intruder mass. Forces at contacts 10 and 25, with the oscillator located at sphere 15, for three values of spring constant, are presented on Fig.~\ref{fig:all_parameter}(a). Transmitted and reflected secondary waves are generated for intermediate values of $k$. These secondary waves are better visualized on insets of Fig.~\ref{fig:all_parameter}(a). They correspond to a succession of delayed nonlinear radiated waves~\cite{Job2007b,Job2009} stemming from --- and thus revealing --- the back and forth oscillations of the resonant intruder. In order to evaluate the relevance of secondary waves on the energy balance, the reflected and transmitted energy, respectively, carried out by these waves are calculated as a function of stiffness, see Fig.~\ref{fig:all_parameter}(b)-(c). A maximum on the reflected waves occurs for a stiffness close to $k=10^5$~N/m, see Fig.~\ref{fig:all_parameter}(b). An optimum of the transmitted energy by secondary waves is also observed for the same value of stiffness, see Fig.~\ref{fig:all_parameter}(c). Together with this maxima, a minimum on the transmitted energy carried out by the main solitary wave is observed, see Fig.~\ref{fig:all_parameter}(c), indicating that the reflected as well as transmitted waves are associated to the same phenomenon. A detailed description of the process of secondary solitary waves generated through the interaction of a incident solitary wave with a local oscillator is beyond the scope of the present work, however, the presented evidence suggest that these waves are generated when two conditions are fulfilled: i) the viscous relaxation time $\tau$ is longer than the oscillation period $T=1/f$, resulting in a weak-to-moderate dissipation, and ii) the resonance frequency $f$ is smaller than the cutoff of the frequency content of the incident solitary wave $\omega_{SW}\propto1/\tau_{SW}$, such that a pulse excites efficiently the resonators and after its passage, they still vibrate and radiate forward and backward secondary waves. In summary, these two conditions read $\tau_{SW}<T<\tau$. Note that the effect of the oscillator mass on the reflected and transmitted energy carried out by secondary solitary waves is presented in Sec.~\ref{sec:appendix_2} of the Appendix; the absorbed energy as well as the amplitude of secondary waves increases with oscillator mass, see Fig.~\ref{fig:sm_fig_2} of the Appendix. The effect of the dissipation of the harmonic oscillator is also investigated and presented in Sec.~\ref{sec:appendix_2} of the Appendix. It is corroborated that generation of secondary waves, reflected and transmitted, depends strongly on the rate of dissipation energy of the harmonic oscillator, see Figs.~\ref{fig:sm_fig_3}(d) and~\ref{fig:sm_fig_3}(e) of the Appendix, respectively.

\begin{figure}[t]
\centering
\includegraphics[width=\columnwidth]{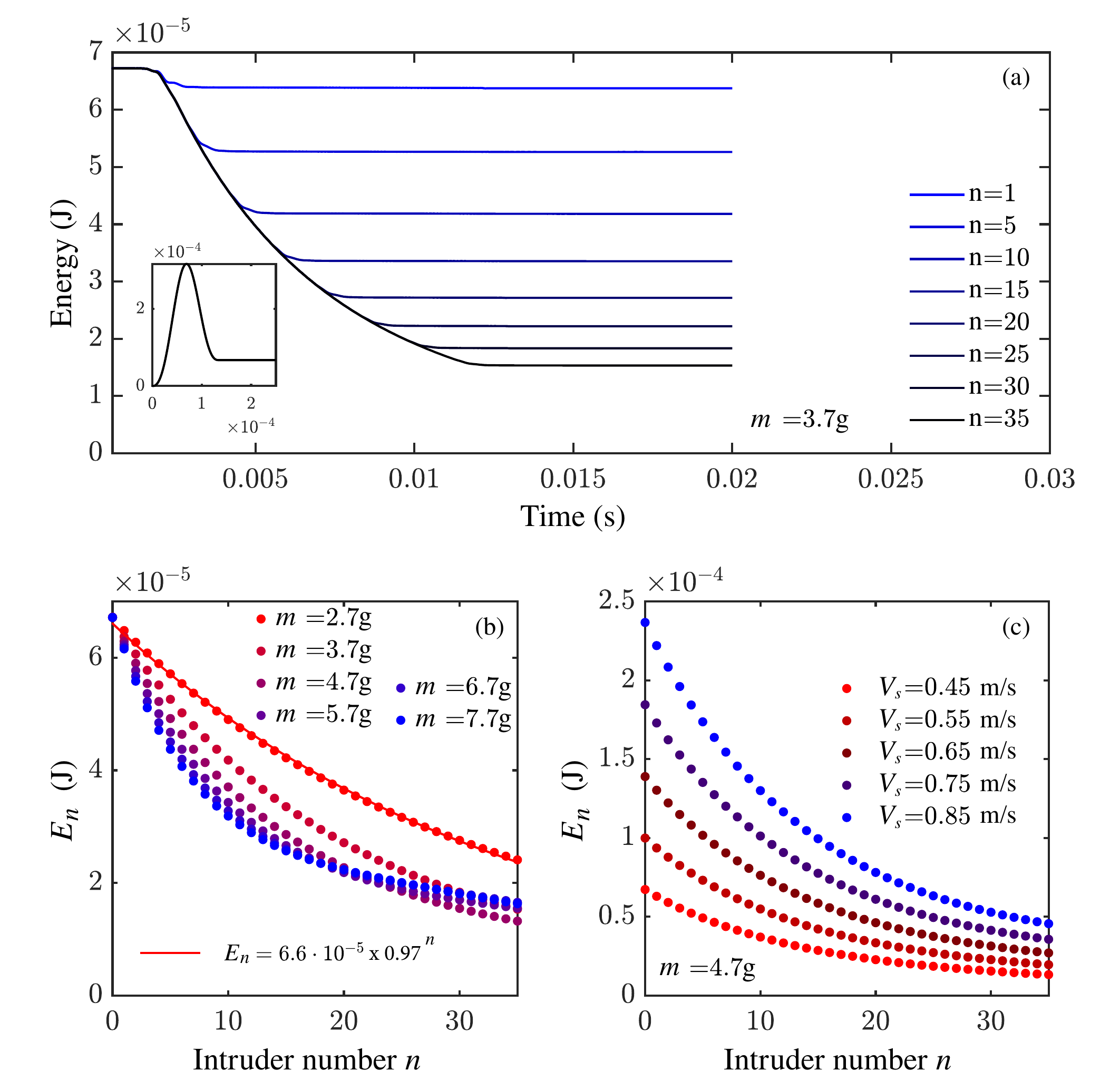}
\caption{(a) Total chain energy versus time, for chains of increasing number, $n$, of intruders. In all cases, the first seven spheres do not contain intruders and sphere 1 is the striker. All chains are excited in the same manner using a steel striker $R_s=4.5$~mm at $V_s=0.45$~m/s. Parameters are $k=10^5$~N/m and $c=12$~N~s/m. (b) Total energy of the chain as a function of intruder number, $n$, with different mass. An exponential decrease of the chain energy with $n$ is observed for small $m \ll M$. However, increasing $m$ leads to an energy absorption process that is more complex and a different functional dependence. (c) Total energy of the chain as a function of intruder number, with mass $m=4.7$~g, for different striker velocity $V_{s}$.}
\label{fig:varios_intrusos}
\end{figure}

%----------%
\subsection{Multiple absorbers}

We now focus on the effects of adding consecutive beads with oscillators to the chain and explore how these oscillators contribute to the absorption of energy from the incident solitary wave. The experimental study of this configuration can be laborious, and since numerical calculations presented previously captures all the experimental observations, we take advantage here to extrapolate these effects from simulations only, in a first attempt. If an incident solitary wave crosses the region of successive oscillators, a certain fraction of the energy carried out by the solitary wave is transmitted forward to the next oscillators, while the rest is reflected backward or absorbed. Figure~\ref{fig:varios_intrusos}(a) shows the effect of energy reduction in chains of spheres with increasing number of oscillators. Under the parameters used, the energy absorbed by an intruder is rapidly dissipated by viscoelasticity.
An exponential decay of the chain energy with $n$ is observed for small $m \ll M$, see Fig.~\ref{fig:varios_intrusos}(b): at each passage of the incident solitary wave through an intruder, one can neglect the reflection thanks to the weak impedance mismatch between the original bead and the one hosting a resonator. This situation can simply be expressed as $E_{n+1}=\mathcal{T} E_n$, where $\mathcal{T}=1-\mathcal{A}-\mathcal{R}$ is the transmission coefficient, with $\mathcal{A}$ the absorption coefficient and $\mathcal{R}\ll1$ the reflection coefficients. In this case, no energy retropropagates backward but only propagates forward, such that $E_{n}=\mathcal{T}^n E_0$, where $E_0$ is the energy carried out by the incident solitary wave prior any intruder interaction. In practice, we find $\mathcal{T}\simeq0.97$ for the case described in Fig.~\ref{fig:varios_intrusos}(b), with the lightest intruder mass, $m=2.7$~g.
Consistently, for increasing $m$, the transmitted solitary wave generates increasing amplitude secondary solitary wave, leading to an energy absorption process that is more complex. Chain energy decreases rapidly with number of oscillators but the functional dependence is no longer exponential, Fig.~\ref{fig:varios_intrusos}(b). 
In fact, as $m$ increases and the mass of host bead decreases, the harmonic oscillator significantly modifies solitary wave propagation. The effect of the presence of heavier oscillators is to induce secondary solitary waves whose amplitude increases as these waves propagate through the chain zone containing local oscillators. Further mass increase leads to the generation and propagation of tertiary solitary waves, see Sec.~\ref{sec:appendix_3} and Fig.~\ref{fig:sm_fig_4} of the Appendix. In this regime, the amplitude of secondary and tertiary waves is increased by the passage through the chain of oscillators, which reveals that the energy transferred from the main solitary wave is radiated in the forward direction; the efficiency of the local oscillator to dissipate the absorbed energy is thus reduced. Note that studying further the mechanisms of amplification of secondary and tertiary waves is beyond the scope of the present work. The effect of multiple successive absorbers is investigated experimentally in Sec.~\ref{sec:appendix_3} of the Appendix and the results are contrasted to numerical calculations. A satisfactory agreement is obtained, see Fig.~\ref{fig:sm_fig_5} of the Appendix, which provides further support to numerical predictions. Finally, the total energy of the chain as a function of intruder number, for distinct striker velocity, $V_{s}$, with fixed mass shows that the process of energy absorption operates robustly over many different striker's input energy, see Fig.~\ref{fig:varios_intrusos}(c).

\begin{figure}[t]
\centering
\includegraphics[width=\columnwidth]{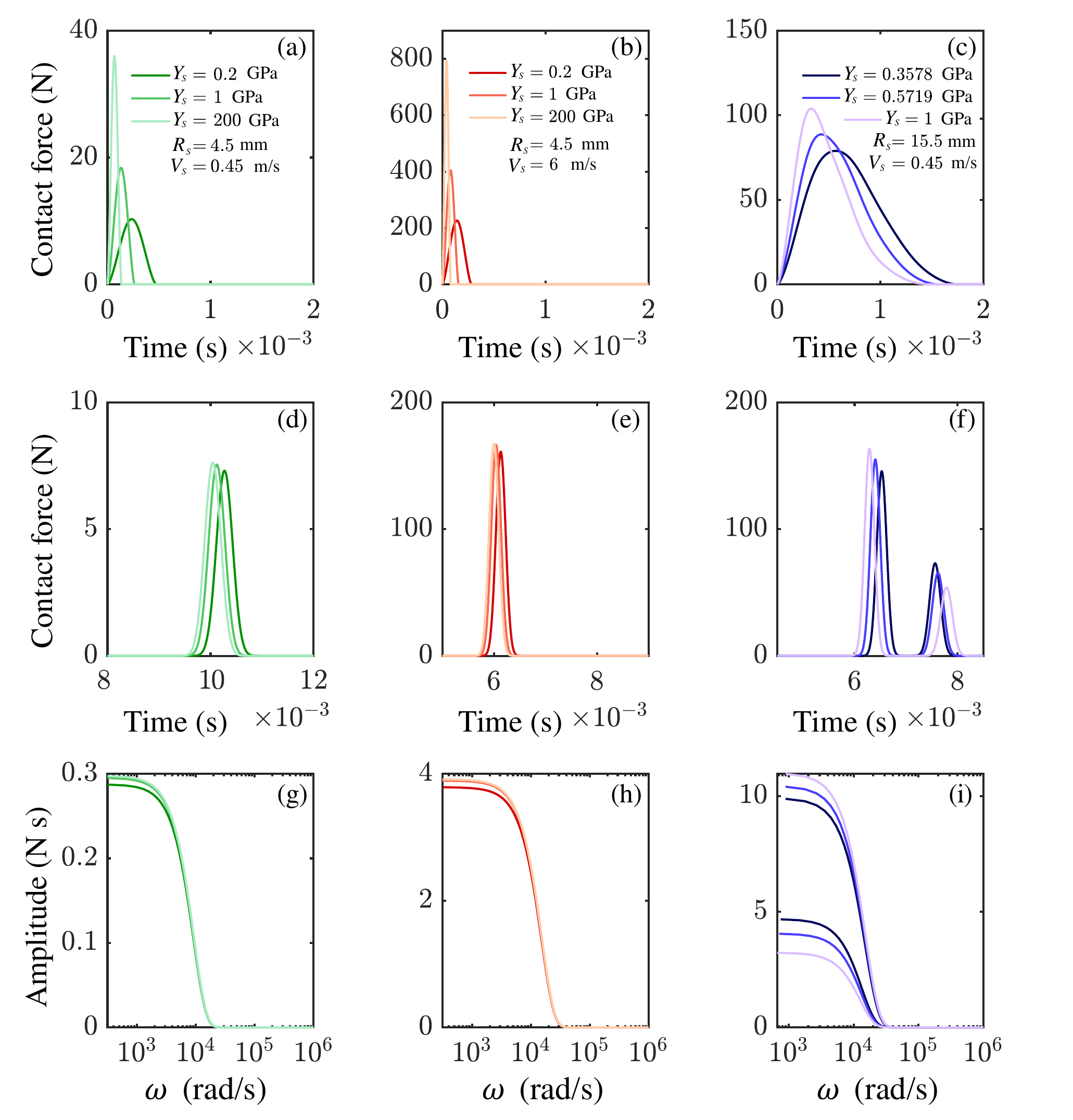}
\caption{(a),(b),(c) Contact force at the striker and the first chain bead for several Young's modulus and representative values of striker size and speed, as indicated at each panel. (d,(e),(f) Contact force resulting from the solitary wave passage at bead 30 for the same parameters as (a),(b),(c). (g),(h),(i) Frequency content of solitary waves, as detected at contact 30. In (i), frequency content of secondary solitary waves are also shown.}
\label{fig:impact_time}
\end{figure}

%----------%
\subsection{Energy injection}

To complete the description we investigate how the energy effectively injected into the chain and how its spectral distribution can be tuned to the frequency response of intruders in order to achieve optimal energy absorption of any excitation. We first explore the effect of the duration of the external impact (impactor-first chain's bead ) on the frequency content of the solitary wave generated. Three parameters influence this impact duration, the Young's modulus, the speed, and the size of the impactor. Using the numerical simulation, we observe that changing the Young's modulus by 3 orders of magnitude produces a factor 3 variation in the duration, see Fig.~\ref{fig:impact_time}(a). Interestingly the solitary wave produced (analyzed at contact 30) does not present significant variations, see Fig.~\ref{fig:impact_time}(d). Frequency content of the propagating solitary wave shows similar trends, see Fig.~\ref{fig:impact_time}(g). 

The effect of striker speed can be visualized through the comparison of Fig.~\ref{fig:impact_time}(a,d,g) to Fig.~\ref{fig:impact_time}(b,e,h), respectively. Although the contact force increases with impacting speed, the impacting time is only reduced by a factor 2. Comparison of Fig.~\ref{fig:impact_time}(d) to Fig.~\ref{fig:impact_time}(e) (solitary waves at contact 30) reflects significant changes on the amplitude of solitary wave with impacting speed. However, the frequency content of solitary waves is only slightly modified, Fig.~\ref{fig:impact_time}(g) to Fig.~\ref{fig:impact_time}(h). Finally, we explore the effect of increasing the striker size. A significant increase of the impact time is observed with striker size, see Fig.~\ref{fig:impact_time}(c), which dramatically affects the solitary wave propagation, see Fig.~\ref{fig:impact_time}(f), leading to the creation of a train of secondary solitary waves. This kind of secondary wave generation has been described in Refs.~\cite{Doney2005,Job2007a,Job2007b}, and its origin associated to a nonlinear mechanism. Again, little effect on the frequency content of the solitary waves is observed, see Fig.~\ref{fig:impact_time}(i).
To provide support to this result, we investigate experimentally how solitary waves of a characteristic frequency distribution develop, see Fig.~\ref{fig:sm_fig_6} of the Appendix.
For stiff impacters, see Fig.~\ref{fig:sm_fig_6}(a) of the Appendix, a single solitary wave is injected into the chain, the  frequency contents does not vary significantly with impacter radius, as predicted by numerical simulations. Interestingly, when the mass of impacter is increased or the stiffness is decreased, see Figs.~\ref{fig:sm_fig_6}(c,d,g) of the Appendix, the collision time increases significantly and, as predicted by numerical calculations, the generation of a train of solitary is indeed experimentally observed.  Comparing the Fourier transform of the transmitted solitary waves, it is corroborated that their frequency content is nearly insensitive to changes in Young's modulus and mass of the impacter, see Figs.~\ref{fig:sm_fig_6}(b,d,f,h) of the Appendix.

\begin{figure}[b]
\centering
\includegraphics[width=\columnwidth]{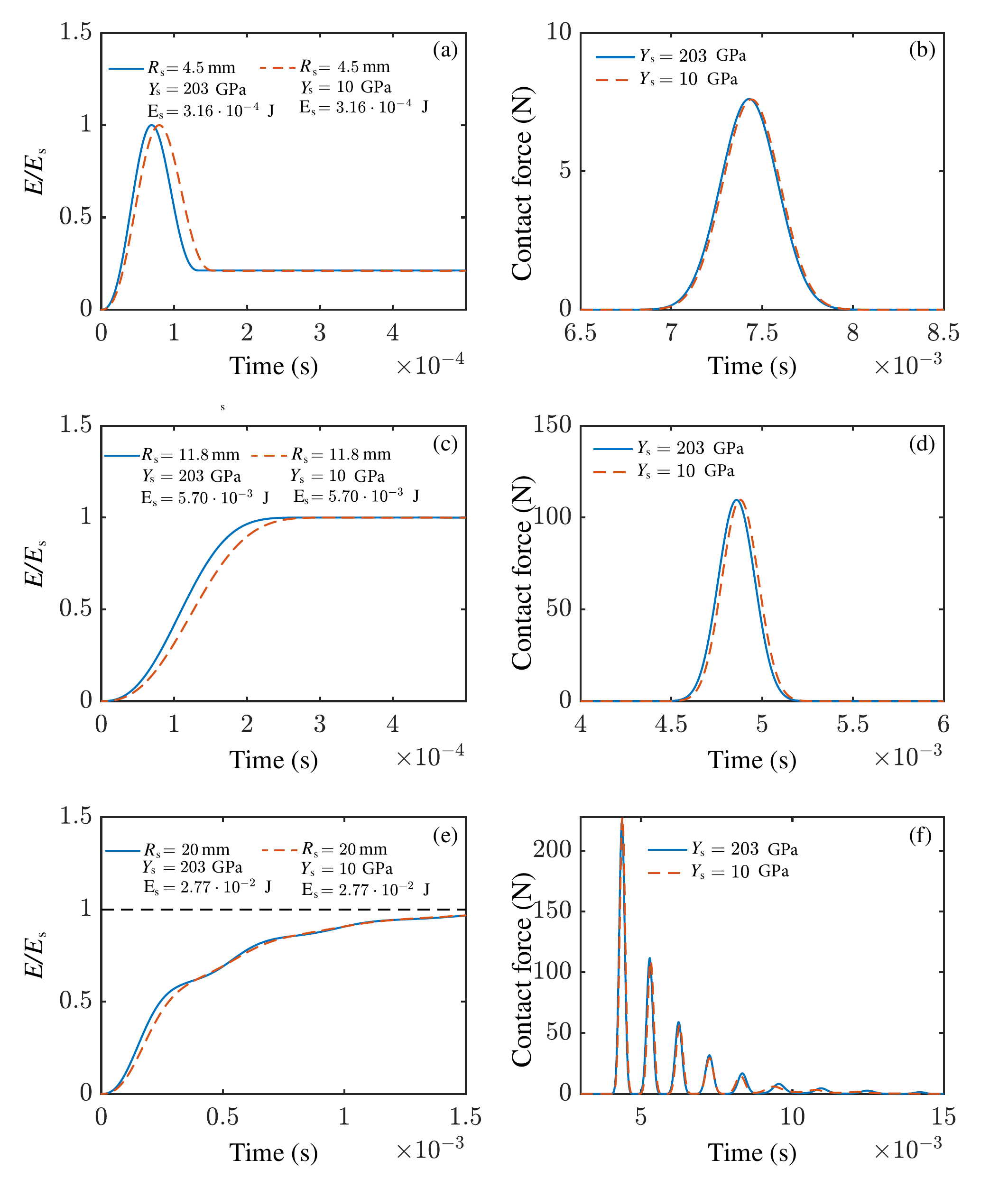}
\caption{Energy of the chain as a function of time. (a) Striker is lighter than a bead of the chain. Energy first increases, it reaches a maximum corresponding to zero kinetic energy of striker and then decreases to reach a constant value. (b) Solitary wave at contact 30 generated under the conditions of (a). (c) Mass striker is close to the mass of chain beads. (d) Solitary wave at contact 30 generated with the impact of (c). (e) Mass striker larger than the mass of chain beads. Energy is totally transmitted to the chain by secondary solitary wave generation. (f) Train of solitary waves generated in (e).}
\label{fig:energy_trans}
\end{figure}

Hereafter, we turn our attention to the energy effectively injected to the chain after striker impact. The main parameter controlling such energy transfer is the ratio of the mass of striker to the mass of a bead. At impact, during compression, the kinetic energy of the striker is gradually transformed into elastic energy at the first bead contact. This energy is then transformed back into kinetic energy, which results in the striker rebound if it is lighter than chain beads, see Fig.~\ref{fig:energy_trans}(a),(b). However, for a mass striker close to that of beads, energy is completely transferred to the chain, see Fig.~\ref{fig:energy_trans}(c),(d). In turn, for $M_s>M$, the energy is totally transmitted to the chain due to the generation of a train of secondary solitary waves, see Fig.~\ref{fig:energy_trans}(e),(f). This process takes longer. 
Further insight about the variables that limit the energy transferred to the chain can be obtained from the analysis of a binary collision model. In this simplified view, the binary collision duration approximates the duration of the propagating solitary waves. Indeed, using energy conservation and the Hertz potential, the collision duration, $t_c$, between two spheres~\cite{Landau2008} and maximum force at collision, $F_m$, read,

\begin{eqnarray}
t_c &=& 2.94 \left(5/4\right)^{2/5} \left(\mu_{1,2}/\kappa_{1,2}\right)^{2/5}V_{1,2}^{-1/5}, \label{eq:col_time}\\
F_m&=&\left(5/4\right)^{7/5}\left ( \mu_{1,2}^{3}\kappa_{1,2}^{2}\right)^{1/5}V_{1,2}^{6/5}, \label{eq:fmax}
\end{eqnarray}

\noindent where $V_{1,2}$ is the collision velocity of bead $1$ on bead $2$, $\mu_{1,2}=(m_1^{-1}+m_2^{-1})^{-1}$ is the reduced mass between masses $m_1$ and $m_2$, and $\kappa_{1,2}$ is defined below Eq.~(\ref{eq:wave_equation}).\\

From Eqs.~(\ref{eq:col_time}) and~(\ref{eq:fmax}) it is seen that the collision duration weakly depends on the striker speed and on the maximum force. By extension to the sphere-sphere collisions in the chain, it explains the weak dependence of the solitary wave duration (or equivalently its frequency content) on the experimental parameters. The condition for maximum momentum and energy transfer from the striker to the first chain bead is easily found using momentum and energy conservation. Thus, after collision, the striker speed $V^{'}_s$ 
and the sphere speed $V^{'}$ become $V^{'}_s= V_s (M_s - M)/(M_s+M)$ and $V^{'}=  2 V_s M_s/(M_s+M)$, from which we deduce that maximum energy transmission occurs for $M_s=M$.

However, total energy transmission also occurs for $M_s > M$. Indeed, in the framework of beads' collision, the generation of secondary waves, which occurs if $M_s > M$, can be interpreted as the generation of more than one collision during the striker impact~\cite{Job2007b}: this is consistent with the fact that the duration of interaction between the striker and the first bead exceeds the collision duration of the two nearest beads of the chain. Thus, a collision of two beads of the chain ends before the interaction of the striker with the first bead of the chain has ended . 

%==========%
\section{\label{sec:conclusions}CONCLUSIONS}

In conclusion, a chain of beads, barely touching one another, in which one bead holds an intruder has been built as a proof of principle of an energy absorber. Numerical solutions of the set of coupled equations quantitatively account for the main features of the absorption process observed experimentally through the tuning of the oscillator dissipation, which allowed for the validation of the mechanical model of the system. 

A generalized systematic numerical study demonstrated that embedded oscillators constitute an efficient manner of pumping the energy from the solitary waves propagating in the chain, leading to an energy absorption that is nearly exponential in the number of oscillators. The analysis shows that in order to obtain optimized energy absorption, the oscillator resonance frequency has to be less than the typical frequency content of the incident solitary wave. In addition, nonlinearities due to Hertz's potential allows for energy absorption in a broad band of excitation frequencies. This property of the chain is due to the fact that the selected solitary wave is dominated by material parameters but it is a weak function of the striker speed. Finally, when the intruder damping is small, generation of secondary solitary waves occurs through radiation. In such a case, the process of energy absorption still holds, but its duration increases with the typical dissipation time.

In this study, we have thus probed the concept of an efficient dissipation in a given frequency band, stemming from the nonlinear conversion of any mechanical excitation. Practically, our laboratory prototype, an alignment of noncohesive spheres including resonant dampers, has two limitations: it has no restoring lateral force preventing the alignments to buckle, and no tensile strength. Therefore, a more concrete implementation requires a rigid guide to align the spheres (e.g. a cylindrical tube) and a compression force to maintain the cohesion (e.g. an internal spring or the own weight of the device to be protected), as proposed in a patented granular chain-based vibration filter~\cite{Boechler2016}. For instance, such a device is considered as a payload vibration isolation system at NASA's {\em Jet Propulsion Laboratory}, in the form of an arrangement of six single-axis alignments into a hexapod~\cite{Dillon2013}. Further potential applications refer to vehicle shock absorbers and earthquake protection systems~\cite{Boechler2016,Dillon2013}. Here, implementing our concept in such devices is likely to provide strong dissipation rate in compact systems, extending their capabilities to harsh mechanical shocks. Furthermore, our proof of concept is not limited to granular-based systems. For instance, metadampers, such as MIM scatterers embedded in an elastic matrix~\cite{Hussein2013}, are also likely to benefit from a nonlinear wave conversion during propagation, prior to resonant dissipation. As another example of concrete implementation, our concept is also transposable to acoustic waves in fluids. Solitary waves being observable in waveguides branched to an array of nonlinear resonators~\cite{Richoux2015}, a nonlinear frequency conversion preceding a tuned absorption of the sound is therefore also conceivable.

%==========%
\section*{ACKNOWLEDGMENTS}
F.M. acknowledges ANID-Chile through Fondecyt Project No. 1201013 and Fondequip No. 130149. Support from LIA-MSD France-Chile (Laboratoire International Associ\'e CNRS, "Mati\`ere: Structure et Dynamique") is greatly acknowledged. We acknowledge the support from DICYT Grant USA1956-042131MH-POSTDOC, Vicerrectoría de Investigación, Desarrollo e Innovación, of Universidad de Santiago de Chile. C.V. acknowledges ANID National Doctoral Program Grant No. 21201036.

%==========%
\begin{appendix}

\renewcommand\thesubsection{A\arabic{subsection}}
\setcounter{subsection}{0}

\renewcommand\theequation{A\arabic{equation}}
\setcounter{equation}{0}

\renewcommand\thefigure{A\arabic{figure}}
\setcounter{figure}{0}

\section*{APPENDIX}

%----------%
\subsection{\label{sec:appendix_1}Frequency response of intruder}

\begin{figure*}[h]
\centering
\includegraphics[width=15cm]{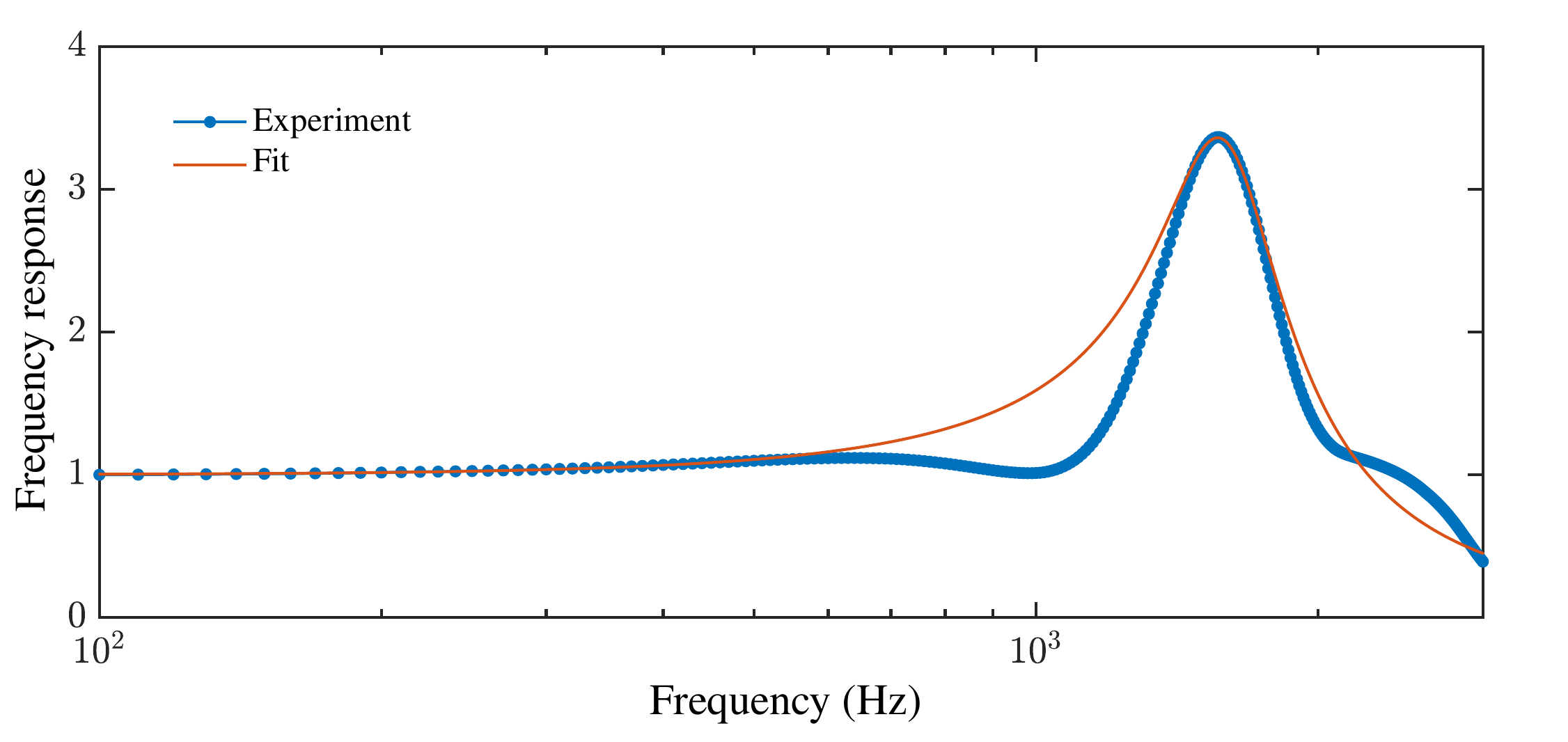}
\caption{Experimental frequency response of intruder obtained by sweeping the excitation frequency. The fit corresponds to the model given by Eq.~\ref{eq:sm_freq_resp}. The fitted resonant frequency is $f_{res} \approx 1600$~Hz and the damping coefficient is $c \approx 12 $~N~s/m.}
\label{fig:sm_fig_1}
\end{figure*}

In order to further characterize the experimental system, and as a check of consistency, the frequency response of the intruder has been measured by exciting a bead embedding an intruder by means of an electromagnetic shaker. An additional accelerometer, glued on the bead, allows for the measurement of the instantaneous displacement $U(t)$ imposed by the shaker to the embedding sphere. As in all other experiments the intruder is a miniature acceleration glued on the polymer layer, which measures its own acceleration, from which one gets its displacement $u$. Experimental results in Fig.~\ref{fig:sm_fig_1} demonstrate an intruder resonant frequency that is $f_{res}=\sqrt{k/m}/2\pi \approx $1600~Hz. In addition, the damping coefficient is found close to $c \approx 12 $~N~s/m. Both values are obtained through a fitting procedure to the experimental data, using the theoretical frequency response of the experimental system, that is, 
\begin{equation}
\tilde{h}(\omega) = \frac{\tilde{u}(\omega)}{\tilde{U}(\omega)} = \frac{k+j\omega c}{k+j\omega c-m\omega^2},
\label{eq:sm_freq_resp}
\end{equation}
where $\tilde{u}(\omega)$ stands for the Fourier transform of $u(t)$ and $m=3.7$~g is the known mass of the inner accelerometer. Both the resonant frequency and the damping coefficient are consistent with the value obtained {\em in-situ}, from the fit of a wave experiment (see the main paper).

%----------%
\subsection{\label{sec:appendix_2}Energy reflection and energy transmission at a bead with intruder}

\begin{figure*}[h]
\centering
\includegraphics[width=15cm]{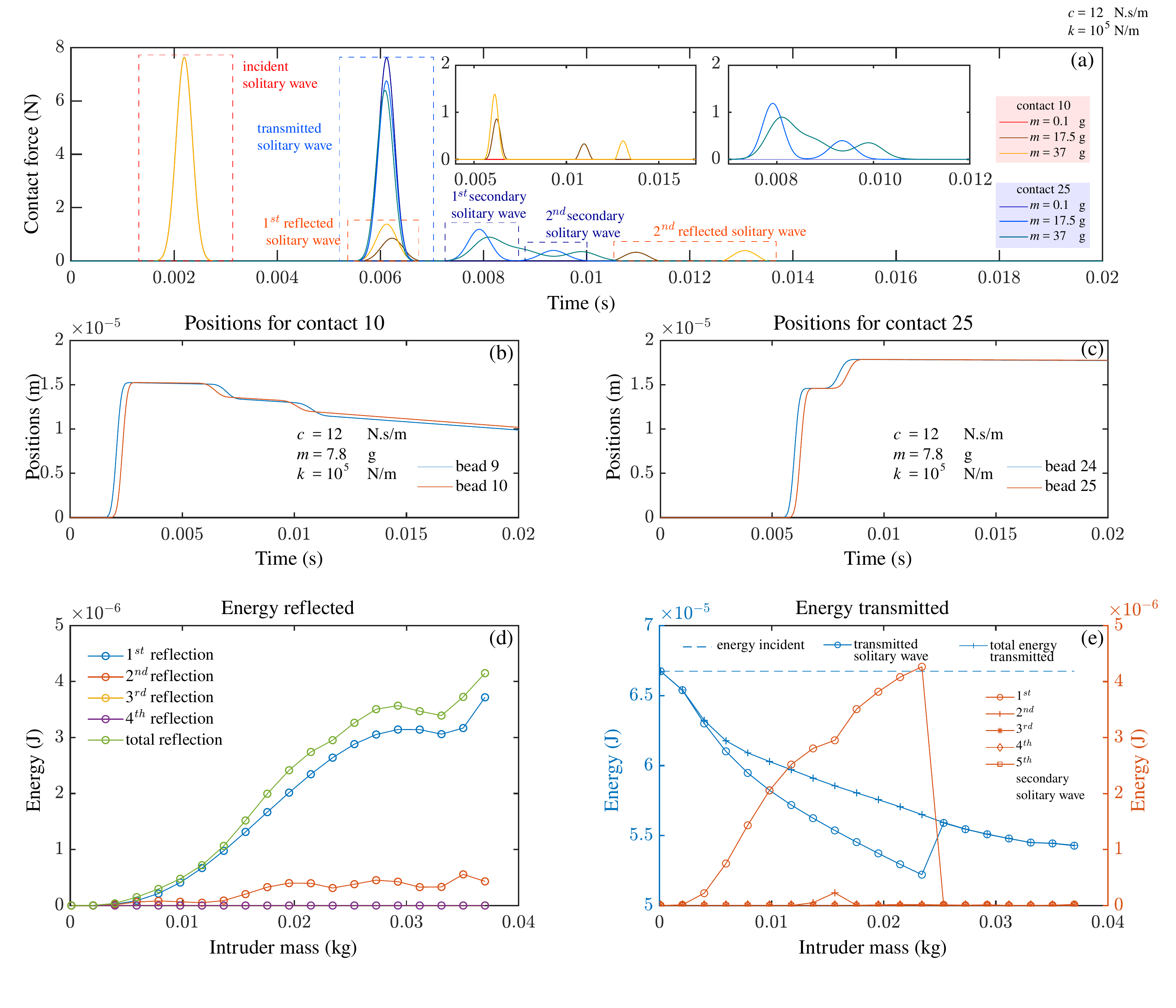}
\caption{(a) Time evolution of forces at contacts 10 and 25, in a chain composed of 46 spheres, illustrating the main and secondary waves after interaction with the oscillator, which is located at sphere 15,  for three values of $m$. Left inset: reflected secondary waves, contact 10. Right inset: transmitted secondary waves, contact 25. (b) Time evolution of positions of beads 9 and 10, illustrating the position shift of beads after interaction, for $m=7.8$~g. (c) Time evolution of positions of beads 24 and 25, illustrating position shift of beads after solitary wave passage, for $m=7.8$~g. (d) Mechanical energy associated with reflected secondary waves as a function of oscillator mass, $m$. (e) Mechanical energy associated with transmitted secondary waves as a function of oscillator mass $m$. For all cases, $c=12$~N~s/m and $k=10^5$~N/m.}
\label{fig:sm_fig_2}
\end{figure*}

The effect of the oscillator mass on the reflected and transmitted energy solitary waves is presented in Fig.~\ref{fig:sm_fig_2}. The force generated by the main transmitted solitary wave together with all secondary waves generated are considered as a function of time, for three distinct values of intruder mass. The absorbed energy at the bead holding the intruder as well as the amplitude of all secondary waves increases with intruder mass, [Fig.~\ref{fig:sm_fig_2}(a)]. The effect on the position of chain beads [Fig.~\ref{fig:sm_fig_2}(b), beads 9-10, and Fig.~\ref{fig:sm_fig_2}(c), beads 24-25] due to the passage of the solitary wave is to introduce a small displacement in the direction of wave propagation, which is seen as positive steps in sphere positions in both figures. Moreover, small jumps in bead position can be seen as the results of passage of secondary waves [Fig.~\ref{fig:sm_fig_2}(b) and~\ref{fig:sm_fig_2}(c)]. Negative jumps in bead position indicating the existence of reflected secondary waves [Fig.~\ref{fig:sm_fig_2}(b)]. These observations are summarized by plotting the reflected energy [Fig.~\ref{fig:sm_fig_2}(d)] and the transmitted energy [Fig.~\ref{fig:sm_fig_2}(e)] as a function of the intruder mass. Thus, energy transmission is reduced with increasing intruder mass. This increase in mass leads also to the increase of amplitude of secondary waves, but these remain small compared to the main incident solitary wave.

\begin{figure*}[h]
\centering
\includegraphics[width=15cm]{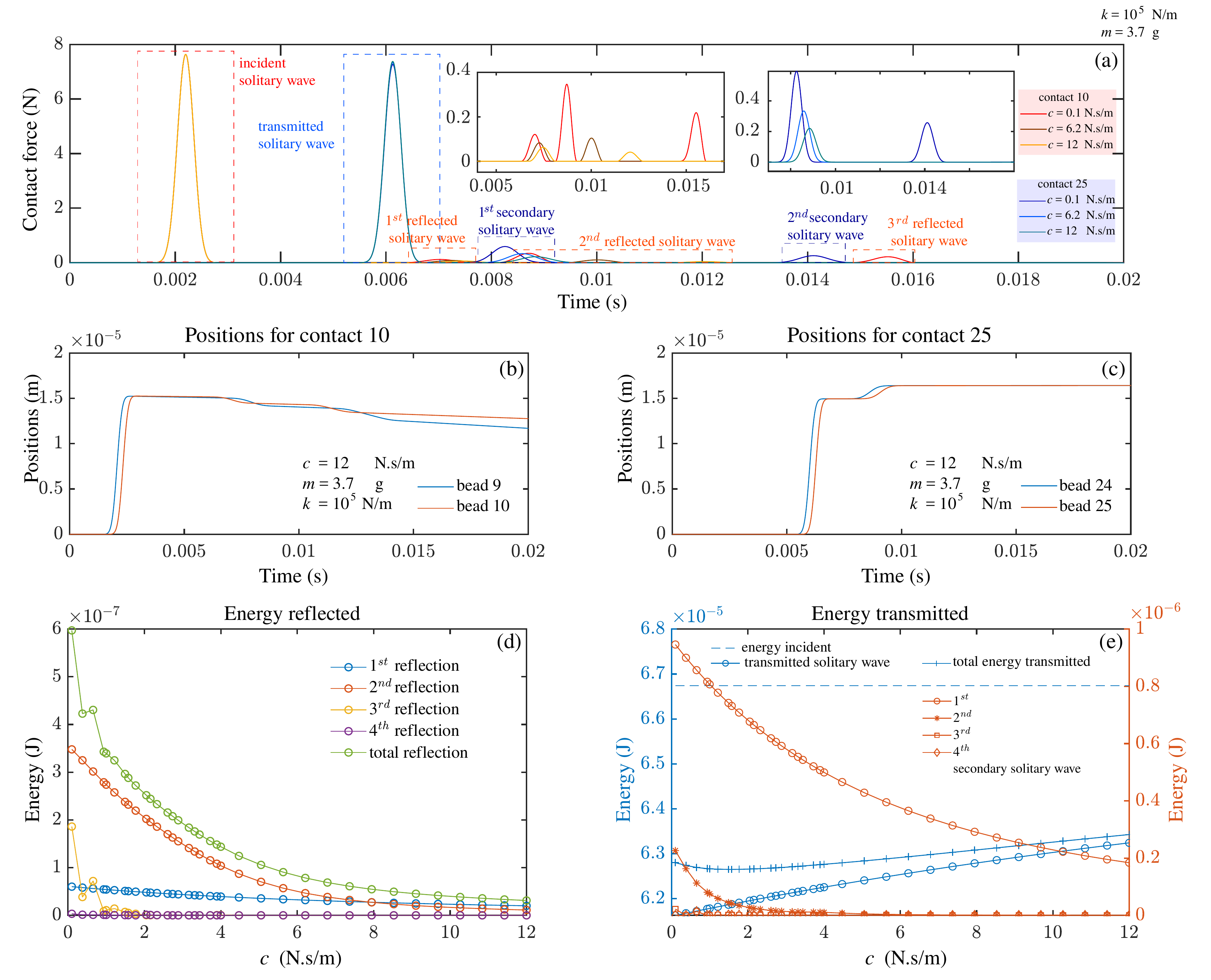}
\caption{(a) Forces versus time, at contacts 10 and 25, illustrating the main and secondary waves after interaction with the oscillator, which is located at sphere 15 (chain composed of 46 spheres) for three values of intruder viscous coupling $c$. Left inset: reflected secondary waves, contact 10. Right inset: transmitted secondary waves, contact 25. (b) Time evolution of positions of beads 9 and 10, illustrating the shift of beads after interaction, for $m=3.7$~g. (c) Time evolution of positions of beads 24 and 25, illustrating the shift of beads after solitary wave passage, for $m=3.7$~g. (d) Mechanical energy associated with reflected secondary waves as a function of oscillator viscous coupling, $c$. (e) Mechanical energy associated with transmitted secondary waves as a function of $c$. For all cases, $m=3.7$~g and $k=10^5$~N/m}
\label{fig:sm_fig_3}
\end{figure*}

The effect of the dissipation of the intruder oscillator on the amplitude of transmitted and reflected waves is also investigated (Fig.~\ref{fig:sm_fig_3}). First, the force generated by the main transmitted solitary and all secondary waves generated during interaction with intruder are considered as a function of time, for three distinct values of viscous coupling, $c$, [Fig.~\ref{fig:sm_fig_3}(a)]. Jumps in bead position are seen as the results of the passage of secondary waves, [Fig.~\ref{fig:sm_fig_3}(b) and~\ref{fig:sm_fig_3}(c)]. It is corroborated that generation of secondary waves, reflected and transmitted, depends  on the rate of dissipation energy of the intruder [Figs.~\ref{fig:sm_fig_3}(d) and~\ref{fig:sm_fig_3}(e)].

%----------%
\subsection{\label{sec:appendix_3}Effect on the solitary wave propagation due to passage through successive heavy intruders}

\begin{figure*}[h]
\centering
\includegraphics[width=15 cm]{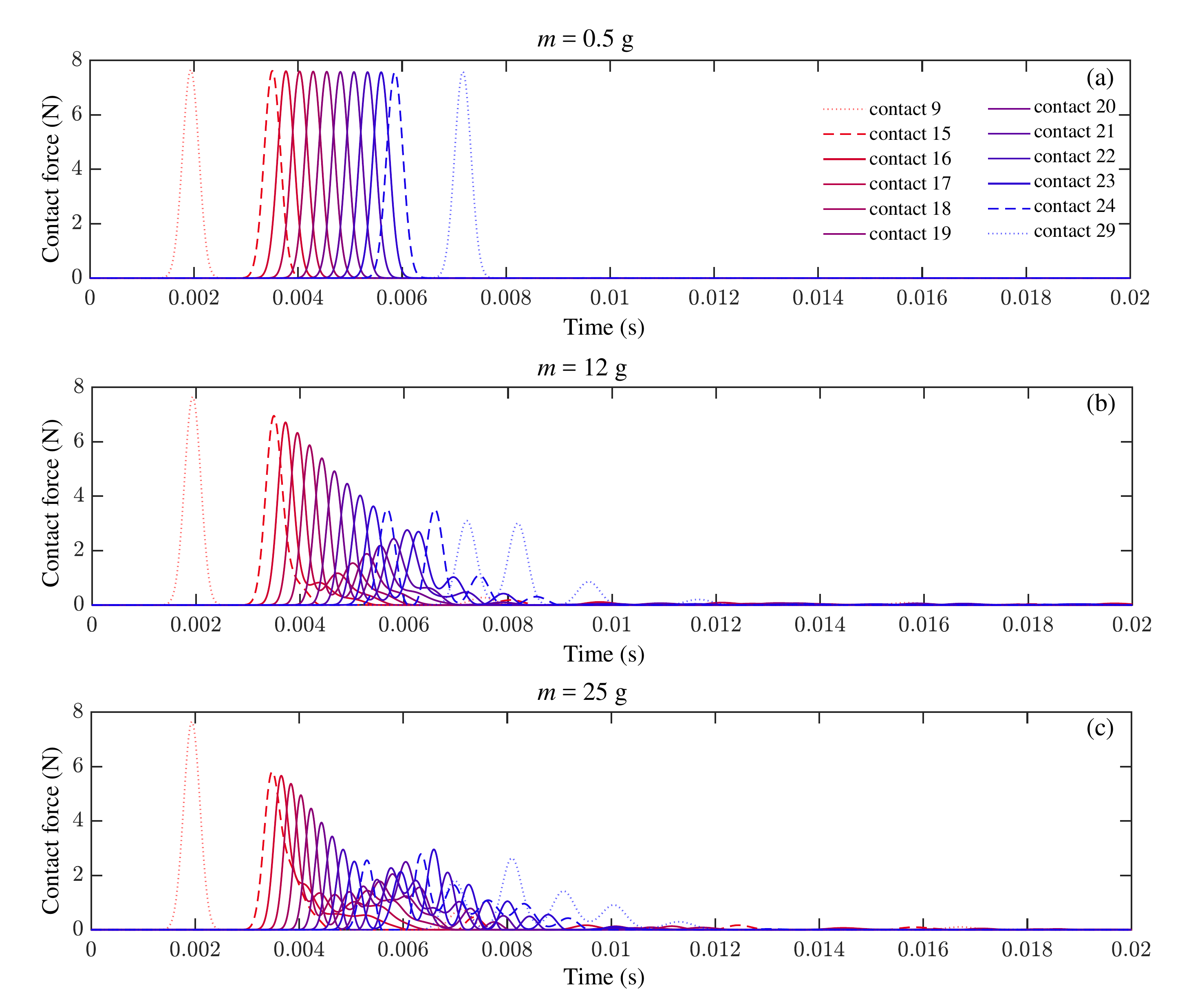}
\caption{Calculated time evolution of solitary wave propagating through an array of 46 spheres, of radius, $R=19$~mm, and mass $51$~g, where nine successive spheres contain equal intruders, starting at sphere number $15$. (a) $m=0.5$~g, (b) $m=12$~g, (c) $m=25$~g.}
\label{fig:sm_fig_4}
\end{figure*}

In the paper, the case of a light intruder has been investigated. In this Appendix, we explore how transmitted and reflected solitary wave are influenced by a heavy intruder. For all cases investigated, the total mass of all beads is kept at a constant value, i.e. if the intruder mass is increased the respective holding sphere mass is decreased in order to keep the sum of both masses equal to the experimental value. The time evolution of solitary wave propagating through an array of 46 spheres, where nine successive spheres contain equal intruders, starting at sphere number $15$ is shown in Fig.~\ref{fig:sm_fig_4}. If the intruder mass is small, it is observed that the amplitude of transmitted solitary wave slowly decreases as it passes through the beads with intruders. As the mass increases, a secondary wave is clearly generated at the first bead with the intruder. The main solitary wave amplitude gradually decreases, while the secondary wave is proportionally amplified by passage through intruders' beads. 

\begin{figure*}[h]
\centering
\includegraphics[width=15 cm]{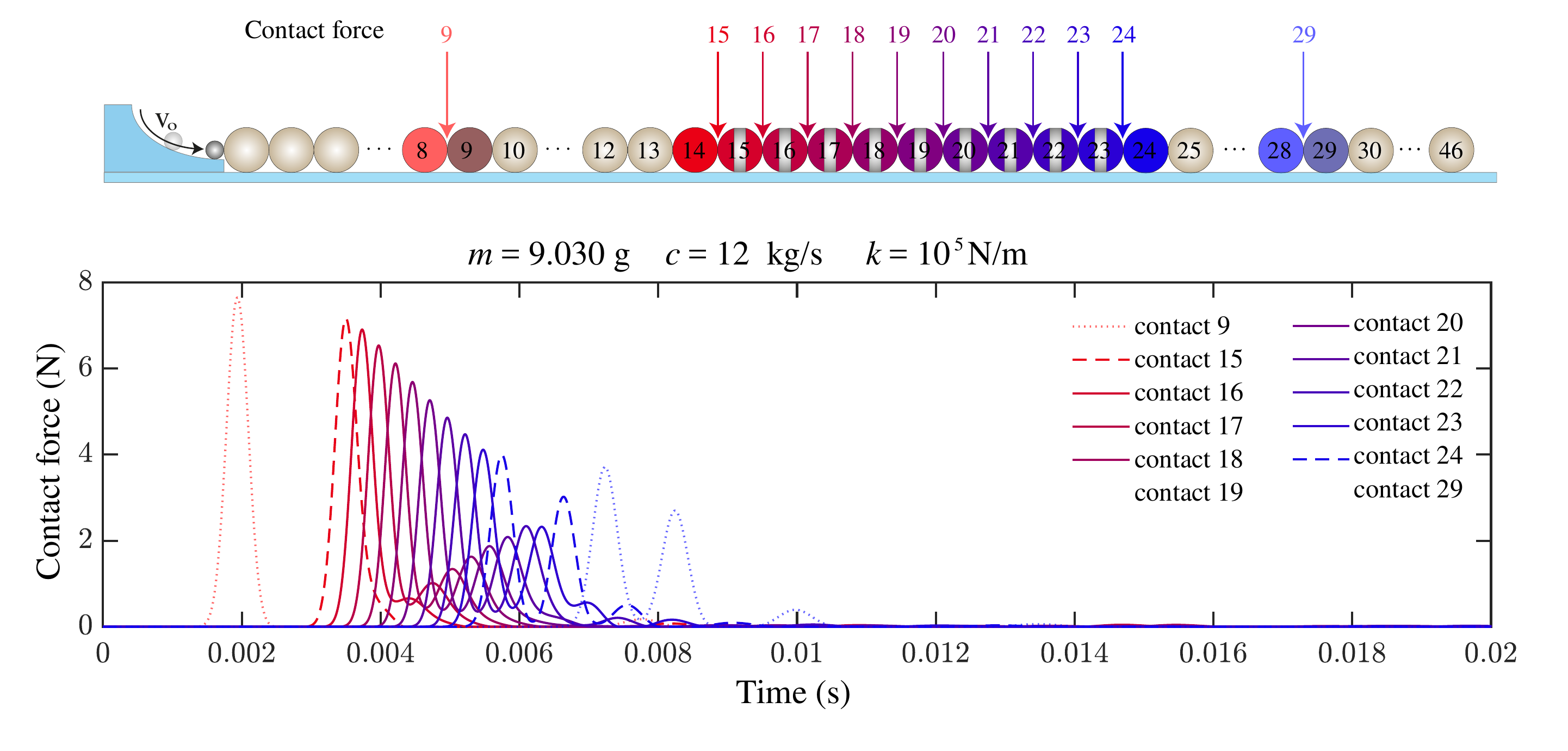}
\caption{Calculated time evolution of solitary wave propagating through an array of 46 spheres, of radius, $R=19$~mm, and mass $51$~g, where nine successive spheres contain equal intruders, starting at sphere number $15$. The figure is a snapshot of an animation, see the Supplemental Material~\cite{VideoSupplMat}, showing the continuous evolution as a function of intruder mass $m$.}
\label{fig:sm_fig_4_animation}
\end{figure*}

For sake of completeness, a detailed variation of solitary wave transmission and reflection with intruder mass is shown in an animation, see~\cite{VideoSupplMat}, whose snapshot is given in Fig.~\ref{fig:sm_fig_4_animation}.

\begin{figure*}[h]
\centering
\includegraphics[width=15 cm]{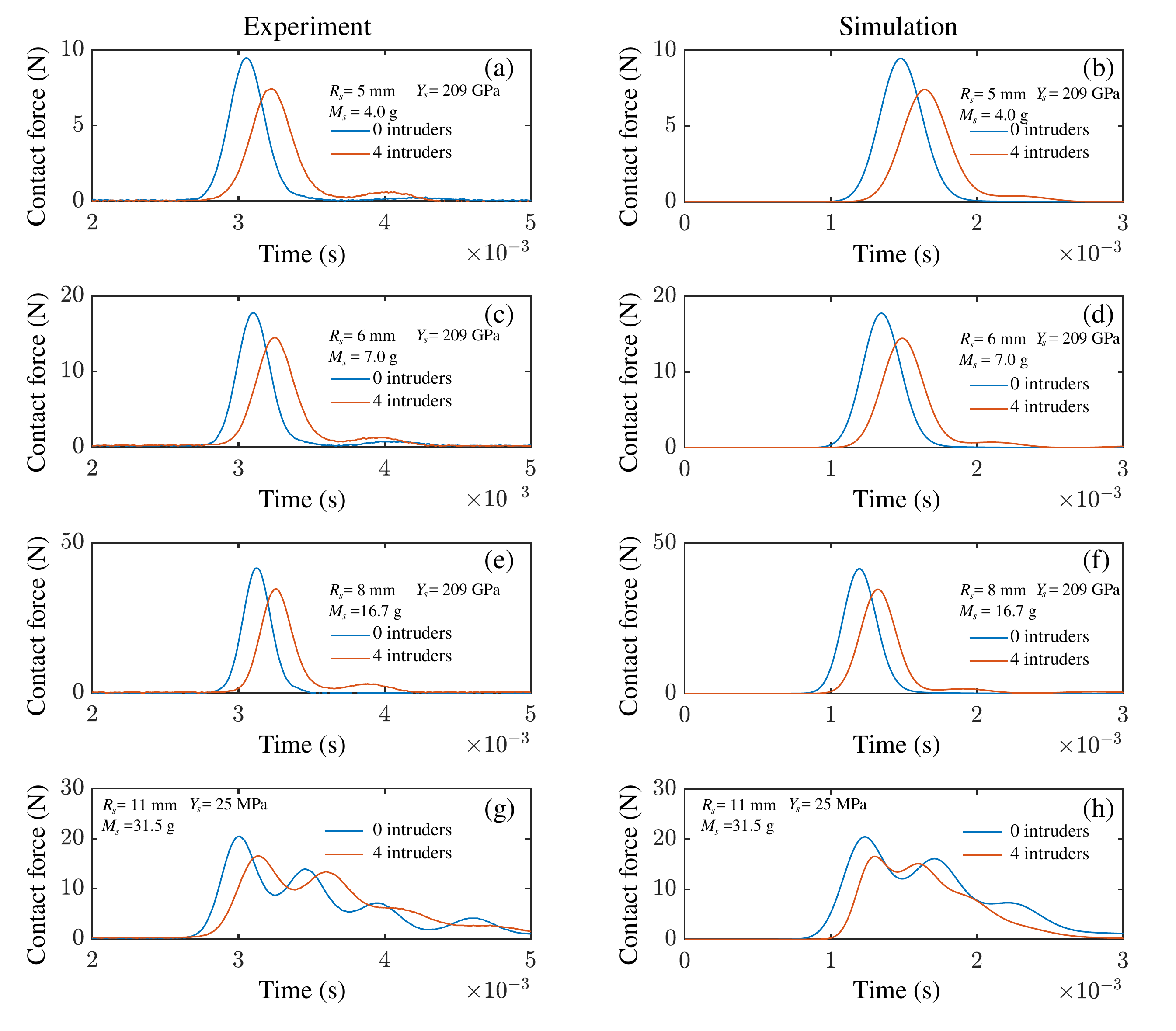}
\caption{Experimental observation of solitary wave propagating through an array of 16 Bakelite spheres, of radius, $R=19$~mm, and mass $51$~g, where four successive spheres contain equal intruders, starting at sphere number $4$, intruder mass,  $m=3.7$~g. In (a),(c),(e), impactors are stainless steel of distinct radius as indicated on each panel; (b),(d),(f) are the numerical predictions under same conditions as (a),(c),(e), respectively. In (g),(h) impactor is made of a soft polymer material. In all cases, signals are considered at contact 12.}
\label{fig:sm_fig_5}
\end{figure*}

To provide experimental proof of the energy absorption features of our system and further consistency with numerical simulation, we build a chain of 16 Bakelite spheres in which a segment,  of four consecutive beads holding each of them an intruder of mass and stiffness equal to that of intruder sensor, is included. Figure~\ref{fig:sm_fig_5} summarizes these experimental results (a), (c), and (e) and compares with numerical simulations (b), (d), and (f), respectively. Using the experimental values of parameters, a satisfactory agreement is provided through numerical simulations.

%----------%
\subsection{\label{sec:appendix_4}Frequency content of chain excitation}

\begin{figure*}[h]
\centering
\includegraphics[width=15 cm]{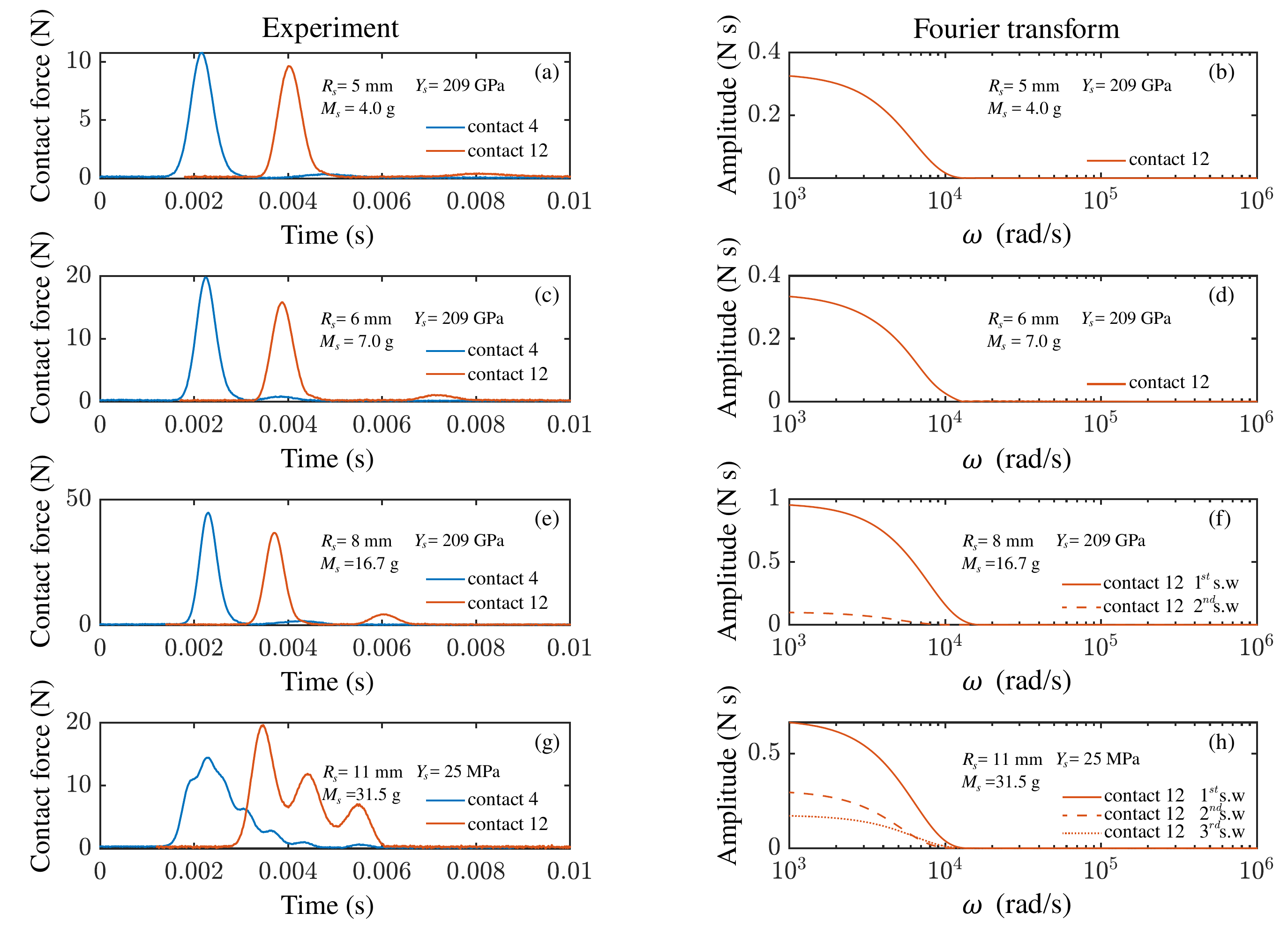}
\caption{Typical solitary wave excitation obtained experimentally at two contacts for stiff impacters of distinct radius (a),(c),(e) and a soft impacter (g). The frequency content of the respective wave excitations is depicted in (b),(d),(f),(h).}
\label{fig:sm_fig_6}
\end{figure*}

To experimentally explore how the chain without intruders distributes the input energy, we impact the chain with distinct impacters and investigate how solitary waves of a characteristic frequency distribution develop. For stiff impacters, a nearly single solitary wave is injected into the chain, Figs.~\ref{fig:sm_fig_6}(a),~\ref{fig:sm_fig_6}(c), and~\ref{fig:sm_fig_6}(e). The respective frequency contents, Figs.~\ref{fig:sm_fig_6}(b),~\ref{fig:sm_fig_6}(d), and~\ref{fig:sm_fig_6}(f), does not vary significantly with impacter radius as predicted by numerical simulations. Interestingly, when the mass of impacter is increased and the stiffness is decreased, the collision time increases significantly and,  as predicted by numerical calculations, the generation of a train of solitary occurs. The separation of the pulses composing the train is more evident as the excitation propagates along the chain, Fig.~\ref{fig:sm_fig_6}g. Once the pulses are separated, each pulse has a width, which has not varied significantly with respect to the previous cases [Figs.~\ref{fig:sm_fig_6}(a),~\ref{fig:sm_fig_6}(c), and~\ref{fig:sm_fig_6}(e)], despite that Young's modulus has been decreased by 4 orders of magnitude. Comparing the Fourier transform of solitary waves, Figs.~\ref{fig:sm_fig_6}(b),~\ref{fig:sm_fig_6}(d),~\ref{fig:sm_fig_6}(f), and~\ref{fig:sm_fig_6}(h), it is corroborated that their frequency content is nearly insensitive to changes in Young's modulus.

%----------%

%----------%

%----------%

%----------%

%----------%

%----------%

%----------%

\end{appendix}

%==========%
\bibliographystyle{unsrt}

\end{document}